\documentclass[11pt]{35}
\usepackage{amsthm, amsmath, amssymb, amsbsy, amsfonts, latexsym, euscript,amscd}
\usepackage{calrsfs} 
\usepackage{graphicx}

\usepackage{hyperref}

\def\le{\leqslant}
\def\leq{\leqslant}
\def\ge{\geqslant}
\def\geq{\geqslant}

\def\phi{\varphi}
\def\hat{\widehat}

\def\kappa{\varkappa}

\theoremstyle{remark}

\numberwithin{equation}{section}

\begin{document}

\noindent{\phantom{DOI}}
\vskip7pt
\noindent

\vskip5mm
\noindent
{\Large \bf
MODELING SKIERS FLOWS
% \\[3pt]
VIA WARDROPE EQUILIBRIUM IN CLOSED CAPACITATED NETWORKS
}

\vskip7mm

{\bf Demyan Yarmoshik}$^*$
\vskip-2pt
{\small Moscow Institute of Physics and Technology}
\vskip-4pt
{\small 9, Institutskiy per., Dolgoprudny 141700, Russia}
\vskip-2pt
{\small
yarmoshik.dv@phystech.edu}

\vskip7pt

{\bf Igor Ignashin}
\vskip-2pt
{\small Moscow Institute of Physics and Technology}
\vskip-4pt
{\small 9, Institutskiy per., Dolgoprudny 141700, Russia}
\vskip-2pt
{\small
ignashin.in@phystech.edu}

\vskip7pt

{\bf Ekaterina Sikacheva}
\vskip-2pt
{\small Moscow Institute of Physics and Technology}
\vskip-4pt
{\small 9, Institutskiy per., Dolgoprudny 141700, Russia}
\vskip-2pt
{\small
sikacheva.ev@phystech.edu}

\vskip7pt

{\bf Alexander Gasnikov}
\vskip-2pt
{\small Moscow Institute of Physics and Technology}
\vskip-4pt
{\small 9, Institutskiy per., Dolgoprudny 141700, Russia}
\vskip-2pt
{\small Innopolis University}
\vskip-4pt
{\small 1, Universitetskaya Str., Innopolis, 420500, Russia}
\vskip-2pt
{\small Institute for Information Transmission Problems}
\vskip-4pt
{\small 19/1, Bolshoy Karetny per., Moscow, 127051, Russia}
\vskip-2pt
{\small
gasnikov@yandex.ru}

\vskip10mm

\parbox{146mm}{\noindent\it
  We propose an equilibrium model of ski resorts where users are assigned to cycles in a closed network.
  As queues form on lifts with limited capacity, we derive an efficient way to find waiting times via convex optimization.
  The equilibrium problem is formulated as a variational inequality, and numerical experiments show that it can be solved using standard algorithms.
}

% Efficiently managing the flow of passengers and skiers in ski resorts is crucial for enhancing customer experience and optimizing operational performance. This study focuses on the problem of reducing waiting times at ski lifts by developing a model that simulates and optimizes the flow of people across the resort's lifts and slopes. By breaking the resort into cycles based on the typical behavior of skiers—where they descend particular slopes and use specific lifts—a graph-based model is proposed to represent the resort's layout. The main objective is to find an equilibrium distribution of skiers, where individuals have no incentive to alter their current choices, leading to an optimized flow. The findings suggest that by adjusting resort parameters such as lift capacities and the distribution of skiers across cycles, significant improvements can be made in reducing congestion. This work offers valuable insights for ski resort management, potentially increasing visitor satisfaction and revenue by decreasing wait times and improving the overall skiing experience.}

{\centering
\section{Introduction}
}

With the increasing popularity of ski resorts, managing the flow of skiers and passengers on lifts has become crucial to ensuring a smooth and enjoyable experience for visitors. The problem lies in optimizing the waiting times at ski lifts, as delays not only frustrate guests but can also affect the overall revenue of the resort. The objective is to create a model that can predict the flow of people across the ski resort, facilitating decision-making for improving client experience and maximizing profit.

The main goal of this study is to develop a model that accurately describes the flow of people across lifts and slopes in a ski resort. 
From the mathematical viewpoint, we uncover fundamental properties of competitive user equilibrium in closed networks with capacitated links.

A large body of literature is devoted to the equilibrium in transportation networks, where each user has a predefined origin-destination pair and is allowed to choose a route to the destination, and possibly, the time of departure or the transportation mode. In this case, the total number of trips is known in advance.
In contrast, in the setup we consider here, the goal of a user is not to commute to a desired destination in minimum time or with minimum generalized cost, but rather to spend some time in the network (without leaving it) in a way to maximize his benefit.
Consequently, since congestion affects travel times on the network's links, the number of trips (e.g., ski descents) for a given user 
% a user makes in a modeled period of time 
depends on the distribution of other users in the network.
% is not an input parameter to the model as it 
To capture this phenomenon, we adopt a simple queuing model and solve for its steady state. The ultimate goal is to find an equilibrium distribution of flows in the network in which no user could change his movement strategy to increase his benefit.
We summarize the main results of our study as follows.
\\
\textbf{Contribution.} 
\begin{itemize}
  \item We propose the first static user-equilibrium model for ski resorts and similar systems;
  \item We provide a novel characterization of the steady state of a closed capacitated network via a system of nonlinear inequalities and show its equivalence to a convex optimization problem;
  \item Through numerical experiments, we observe benign computational properties of the equilibrium user assignment problem.
\end{itemize}
\textbf{Paper organization.} 
In Section~\ref{sec:relwork}, we provide a literature review of ski resorts modeling and equilibrium user assignment in networks.
Section~\ref{sec:main} contains the main part of the work, including the mathematical formulation of our model and analysis of queues.
Results of numerical experiments are presented in Section~\ref{sec:exps}.
Future work is discussed in Section~\ref{sec:discussion}, and concluding remarks are given in Section~\ref{sec:conclusion}.

{\centering
  \section{Related work}\label{sec:relwork}
}

The de facto standard approach for the similar problem of modeling transportation flows in urban networks is Beckmann's equilibrium traffic assignment model \cite{beckmann1956studies}.
This formulation has been mathematically well-developed through decades of research, with numerous extensions addressing multi-class users \cite{decea2005solving}, dynamic models \cite{vickrey1969congestion}, and combined mode-choice scenarios \cite{kubentayeva2024primal}. 
The model's convex optimization formulation enables efficient solution methods ranging from the Frank-Wolfe algorithm to modern column-generation \cite{babazadeh2015reduced} and primal-dual approaches \cite{kubentayeva2024primal}.
Alternative formulations like Nesterov-de Palma's model \cite{nesterov2003stationary} provide complementary perspectives, as thoroughly compared by \cite{chudak2007static}, and can be even more effectively solved using multicommodity flows algorithms.
%%for LLM: add a brief review of how well mathematically developed and widely applied equilibrium traffic assignment (Beckmann) models are

For ski resorts, several studies have examined the movement and behavior of skiers, addressing queue formation, skier flows, and infrastructure optimization.
Most of the works devoted to modeling ski resorts propose multi-agent simulation models.
\cite{pullman2003strategies} used a multiperiod event-based simulation model with a multinomial logit discrete choice submodel to assess management options for improving the profitability of a real ski resort.
They compared 5 different strategies such as changing price policies, customer class variation through marketing, deployment of information services, and capacity expansion, concluding that information signage is the most effective option.
In \cite{poulhes2017dynaski}, a similar Dynaski model, which considers a group of skiers as a single agent, was implemented for a case study at La Plagne resort to evaluate the impact of an old chairlift replacement.
\cite{chetouane2011modelling} also used a discrete-event model to evaluate the solutions for the lift selection problem produced by an integer linear programming model. 
The study \cite{tino2014queues} compares RFID data from Verbier resort with results of a multi-agent simulation to study the formation of queues on lifts and crowding on ski runs.
An extremely minimalist simulation model was used in \cite{biagi2013network} to simulate the effect of disruptions such as lift or piste closure.

As an alternative to multi-agent simulations, \cite{siberg2023are} represents a ski resort as a Jackson network; however, in their model, the probabilities of choosing a lift are fixed, so the users are unable to adapt their strategies to observed waiting times and no competitive interaction occurs.

Some other studies emphasize particular elements of ski resort systems.
\cite{delibasic2017investigation} applied recommendation system algorithms to predict lift choices.
\cite{holleczek2012particle} modeled the movement of skiers on ski slopes by representing them as mass particles affected by social and physical forces, validating the model with GPS traces.

To the best of our knowledge, no equilibrium models for closed networks similar to ski resorts have been suggested in the literature.
The most relevant work here is \cite{brooks2016}, where a Wardrop-like equilibrium principle for closed networks was formulated, and methods for finding system-optimum user allocation were suggested along with a price-allocation policy inducing effective equilibrium.
The main limitation of that work is that it only considers bipartite graphs, while a substantial part of our work is devoted to the analysis of queues in arbitrary directed graphs.

{\centering
  \section{Novel equilibrium model}\label{sec:main}
}

\subsection{Equilibrium and utility}

The resort is represented as an oriented graph whose links are slopes and lifts 
(we assume that any other link of any other type, e.g. horizontal transition between ropeway stations, can be modeled as either a lift or a slope).
Similar to traffic assignment equilibrium models, where user's strategy is a path between origin and destination \cite{beckmann1956studies,nesterov2003stationary,decea2005solving}, skier's strategy in our model is a cycle in the resort's graph.
To define the strategy set we specify all cycles corresponding to the skier's possible motion around the resort; by default we enumerate all cycles in the graph.

To define utility function of a cycle, we need to introduce the following parameters.

\textbf{Uplift.}
The uplift $u \in U$ is characterized by:
\begin{itemize}
    \item Its capacity $b_u$ measured in persons per unit of time. 
    We assume that lift capacities $b \in \mathbb{R}^{|U|}_+$ are normalized by the total number of skiers in the system.
      Similarly to \cite{nesterov2003stationary,pullman2003strategies,poulhes2017dynaski}, we use a simple deterministic point queue model, so that the flow on the lift $f_{u}$ depends on the number of users in the queue $q_u$ and the input flow on the lift $f^{\text{in}}_{u}$ as
    \begin{equation}
      \begin{cases}
        f_u = \min\{b_u, f^{\text{in}}_{u}\}, & q_u = 0,\\
        f_u = b_u, & q_u > 0,
      \end{cases}
    \end{equation}
    and $f^{\text{in}}_u - f_u$ is the queue growth rate.
    We denote by $t_u$ the waiting time a user spends in the queue: $t_u = \begin{cases}
      \frac{q_u}{b_u}, & q_u > 0, \\
      0, & q_u = 0. 
    \end{cases}$ 
    %(in case of a single-elevator system with the lift $i$) or $\displaystyle c=\min_{i \in Lifts}\{c_i\}$ (in general case of a system),
\end{itemize}

\textbf{Slope.}
Each slope (piste) $s \in S$ is associated with a value/worth $v_s$ that measures the entertainment/practice value of skiing the slope estimated by the average skier. 
In general, it seems very reasonable to consider a multi-class formulation to assign different values of attractiveness to slopes for different groups of people, since novice skiers are much less likely to choose expert-level slopes.
Our model can be easily extended to this setup, so we consider only a single user class for clarity.

\textbf{Cycle's utility}.
Value $v_c$ of a cycle $c \in C$ is the sum of value coefficients of all slopes in the cycle $v_c = \sum_{s \in c} v_s$.
We derive our model from the principle that a user aims to maximize total value of visited slopes within a given period of time. 
Equivalently, a skier choose a cycle $c$ with maximum ratio of value per passage time, which we call the utility $\tau_c$ of the cycle:
\begin{equation}
  \tau_c(\mathbf n) = \frac{v_c}{t_c(\mathbf n) + \hat t_c},
\end{equation}
where $\hat t_c$ is the queue-free time to complete the cycle, $\mathbf n$ is the distribution of skiers over cycles and $t_c(\mathbf n)$ is the total waiting time in queues on the cycle: $t_c(\mathbf n) = \sum_{u \in c}t_u(\mathbf n)$.
Since lift capacities are normalized by the total number of skiers, $\mathbf n$ is also normalized and belongs to the unit simplex  $\mathbf n \in \Delta^C = \{\mathbf{n} \in \mathbb{R}^{|C|}~|~ \sum_{c \in C} n_c = 1, ~n_c \geq 0\}$.

In other words, in equilibrium state, all users occupy cycles whose utility is equal to the maximum utility among all cycles.
Cycles with smaller utility are not used. 
The same adaptation of Wardrop's equilibrium principle for closed loop systems was also formulated in \cite{brooks2016} in context of debris removal.
However, they considered only bipartite graphs, where waiting times does not demonstrate such complex dependence on the distribution of users between cycles as in the case of arbitrary directed graphs.
We address the problem of calculation of the waiting times for a given distribution of users in Subsections \ref{sec:queue_system} and \ref{sec:queue_opt}.

\subsection{Optimization problem and competitive equilibrium}

According to Wardrop's principle, a transportation system, after sufficient time, reaches an equilibrium state in which the utility function is equal across all cycles. This competitive equilibrium corresponds to a stable distribution of traffic flows represented by a vector $\mathbf{n}^* \in \Delta^C$, where $\Delta^C$ is the unit simplex. The main objective is to determine this competitive equilibrium distribution. The related work \cite{brooks2016} investigates equilibria on cycles within bipartite graphs, providing foundational insights relevant to our study.

To numerically solve for the equilibrium, we reformulate the problem as a variational inequality, which is standard for traffic assignment \cite{dafermos1980}. The equilibrium distribution $\mathbf{n}^*$ satisfies the following variational inequality 
\begin{equation}\label{eq:vi}
  \langle \boldsymbol{\tau}(\mathbf{n}^*), \mathbf{n} - \mathbf{n}^* \rangle \leq 0, \quad \forall \mathbf{n} \in \Delta^C,
\end{equation}
where $\langle \cdot, \cdot \rangle$ is the standard inner product in the Euclidean space.
%and $\Delta^C = \{\mathbf{n} \in \mathbb{R}^{|C|}~|~ \sum_{c \in C} n_c = 1, ~n_c \geq 0\}$ is the unit simplex.
By definition, any solution to this inequality characterizes a competitive equilibrium distribution that balances the utilities across all cycles, ensuring no individual user can increase his benefit by unilaterally changing his strategy, i.e., changing the cycle.

\subsection{Queue network as nonlinear system of inequalities}\label{sec:queue_system}
We want to find waiting times on lifts $\mathbf{t} \in \mathbb{R}^{|U|}_+$ given the distribution of skiers among the cycles $\mathbf{n} \in \Delta^C$.
In this subsection we require that cycles with no skiers on them ($n_c = 0$) are temporarily excluded from $C$, so that all components of $\mathbf n$ are greater than zero.

Let $f^C \in \mathbb{R}^{|C|}_+$ and $f^U \in \mathbb{R}^{|U|}_+$ denote the auxiliary variables for flows on cycles and lifts respectively,
and $\hat t \in \mathbb{R}^{|C|}_+$ denote queue-free cycle passage times.
Finally, let the matrix $\mathbf \Theta$ encode the relation between lifts and cycles: $\mathbf \Theta_{cu} = 1$ if and only if lift $u$ belongs to cycle $c$.

For steady-state flow we have the following nonlinear system of equations and inequalities:
\begin{subequations} \label{eq:queue_system}
  \begin{align}
  \hat{\mathbf{t}} + \mathbf \Theta \mathbf{t} &= \frac{\mathbf{n}}{\mathbf{f}^C} \label{eq:time_to_flow}
  \\
  \mathbf \Theta^\top  \mathbf{f}^C &= \mathbf{f}^U \label{eq:lift_flow}
  \\
  \mathbf{t} \odot (\mathbf{f}^U - \mathbf{b}) &= \mathbf{0} \label{eq:compl}
  \\
  \mathbf{f}^U &\le \mathbf{b} \label{eq:cap}
  \\
  \mathbf{f}^C \ge 0, \quad \mathbf{t} &\ge 0, \label{eq:nonneg}
\end{align}
\end{subequations}
where the symbol $\odot$ denotes the element-wise (Hadamard) product.
The system has the following meaning
\begin{itemize}
  \item equation \eqref{eq:time_to_flow} states that the number of skiers on a cycle is equal to the flow on it multiplied by passage time (including time in queues on all lifts in the cycle);
\item equation \eqref{eq:lift_flow} defines the flow on a lift (in steady state inflow is equal to outflow) as the sum of flows on the cycles containing the lift;
\item equation \eqref{eq:compl} ensures, according to our queue model, that for each lift either the time in the queue is equal to zero, or the flow is equal to the capacity;
\item inequalities \eqref{eq:cap} are the lift capacity constraints;
\item and \eqref{eq:nonneg} are nonnegativity constraints ($\mathbf f^U \geq 0$ is implied by \eqref{eq:lift_flow}).
\end{itemize}

\subsection{Toy networks}
\begin{figure}[htpb]
  \centering
  \includegraphics[width=0.2\textwidth]{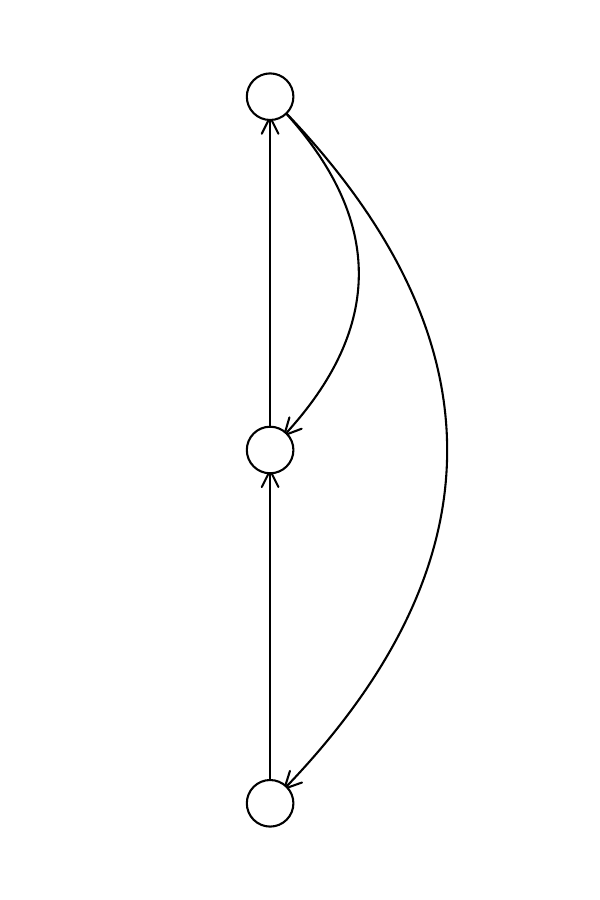}
  \caption{Toy network with two sequential lifts and two slopes}
  \label{fig:2lifts}
\end{figure}

In this section we illustrate the queue network behavior with two toy networks.
\medskip
\\
\textbf{Two lifts.}
Consider a network with two lifts and two slopes illustrated in Figure~\ref{fig:2lifts}.
Let $t_1$ and $t_2$ be waiting times on the lower and the higher lift respectively.
Assign index $c=1$ to the larger cycle containing two lifts and index $c=2$ to the smaller (upper) cycle.
Thus \eqref{eq:lift_flow} becomes $f^U_1 = f^C_1$ and $f^U_2 = f^C_1 + f^C_2$.
We can solve the system \eqref{eq:queue_system} analytically by considering four cases
\begin{itemize}
  \item $t_1 = 0$, $t_2 = 0$. From \eqref{eq:time_to_flow} we immediately get
    $\mathbf{f}^C = \frac{\mathbf{n}}{\hat{\mathbf{t}}}$.
  \item $t_1 > 0$, $t_2 = 0$. Since there is no queue on the upper lift, \eqref{eq:time_to_flow} gives $f^U_2 = \frac{n_2}{\hat{t}_2}$. By \eqref{eq:compl} nonzero waiting time on the first lift implies $f^U_1 = b_1$,  therefore, by \eqref{eq:time_to_flow}, $t_1 = \frac{n_1}{b_1} - \hat{t}_1$.
  \item $t_1 > 0$, $t_2 > 0$. In this case both lift capacities are saturated, so $f^C_1 = b_1$ and $f^C_2 = b_2 - f^C_1 = b_2 - b_1$. 
    Solving \eqref{eq:time_to_flow} yields $t_1 = \frac{n_1}{b_1} - \frac{n_2}{b_2 - b_1} - \hat{t}_1 + \hat{t}_2$, and $t_2 =\frac{n_2}{b_2 - b_1} - \hat{t}_2$.
  \item $t_1 = 0$, $t_2 > 0$. Using that the flow on the upper lift is equal to its capacity with \eqref{eq:time_to_flow}, we have two equations on $f^C_1$ and $t_2$:
    $f^C_2 = b_2 - f^C_1 = \frac{n_2}{t_2 + \hat{t}_2}$, $f_1^C = \frac{n_1}{t_2 + \hat{t}_1}$. 
    Substituting $f_1^C$, we obtain $b_2 - \frac{n_1}{t_2 + \hat{t}_1} = \frac{n_2}{t_2 + \hat{t}_2}$.
    Since denominators are greater than zero, this is equivalent to a quadratic equation, whose only nonzero root (we prove uniqueness of cycle waiting times in the following subsection) is $t_2$. 
\end{itemize}

The solution to the system can be found by considering solutions for all four cases and selecting the feasible one. 
This example shows that even such very simple network demonstrates complex behavior with a nonsmooth, nonlinear, nonconvex/nonconcave dependence of cycle waiting times as functions of user distribution.

\textbf{Parallel slopes.}
\begin{figure}[htpb]
  \centering
  \includegraphics[width=0.2\textwidth]{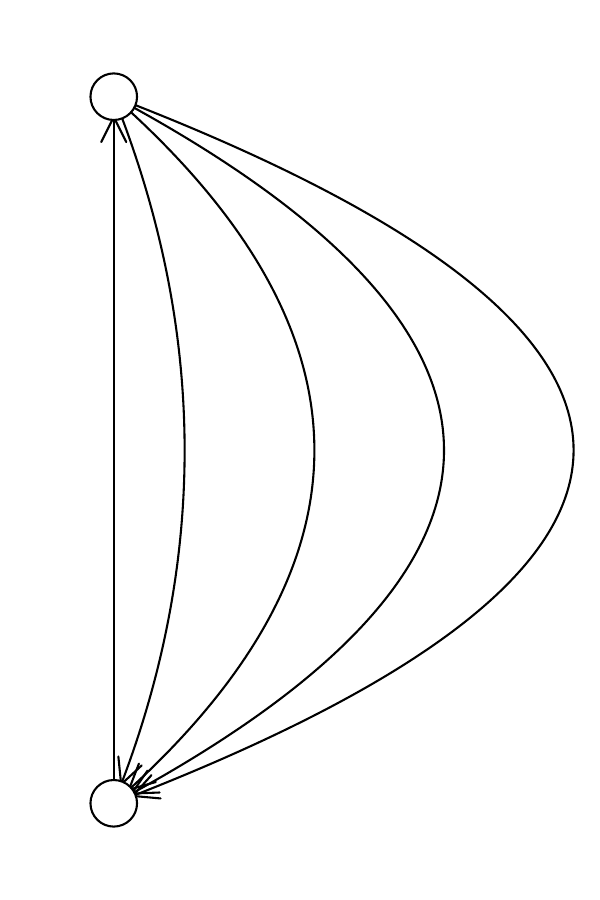}
  \caption{Toy network with parallel slopes}
  \label{fig:par_slopes}
\end{figure}

An extremely simple type of network is shown in Figure~\ref{fig:par_slopes}.
It consists of a single lift and $k$ parallel slopes.
In the presence of queue we can derive cycle flows from the waiting time $t$ using \eqref{eq:time_to_flow}: $\hat{t}_c + t = \frac{n_c}{f^C_c}~\forall c=1,\ldots,k$.
Substituting in into $f^C_1 + \ldots + f^C_k = b$ gives $\frac{n_1}{\hat{t} + t_1} + \ldots + \frac{n_k}{\hat{t} + t_k} = b$ which is equivalent to a polynomial equation of order $k$.

As a result, for this example one cannot even obtain the solution in analytical form. 
To find the waiting time numerically one can apply a linear search algorithm such as bisection or Newton's method.

\subsection{Computation of waiting times and cycle flows given a distribution of users}\label{sec:queue_opt}
Having established the equilibrium conditions as a variational inequality, it is necessary to explicitly determine the cycle waiting times in queues for any given distribution $\mathbf{n}$. This step is essential because the cycle utilities $\mathbf{\boldsymbol \tau}(\mathbf{n})$ depend explicitly on this quantity.

The problem of solving \eqref{eq:queue_system} can be equivalently expressed as finding the optimal primal-dual pair of the following convex optimization problem
\begin{equation}\label{prob:queue_opt}
\min_{\mathbf{f}^C \geq 0} \quad \langle \hat{\mathbf{t}}, \mathbf{f}^C \rangle - \langle \mathbf{n}, \ln \mathbf{f}^C \rangle \quad \text{subject to} \quad \mathbf \Theta^\top \mathbf{f}^C \leq \mathbf{b},
\end{equation}
where the logarithm is applied component-wise to $\mathbf{f}^C$.

Indeed, consider the Lagrangian
\[
\mathcal{L}(\mathbf{f}^C, \mathbf{t}) = \langle \hat{\mathbf{t}}, \mathbf{f}^C \rangle - \langle \mathbf{n}, \ln \mathbf{f}^C \rangle + \langle \mathbf{t}, \mathbf \Theta^\top \mathbf{f}^C - \mathbf{b} \rangle,
\]
where the dual variables $\mathbf{t} \geq 0$ correspond to the capacity constraints. 
Writing down the Karush–Kuhn–Tucker (KKT) conditions given by the primal optimality condition
\begin{equation*}
\nabla_{\mathbf{f}^C} \mathcal{L} = \hat{\mathbf{t}} - \frac{\mathbf{n}}{\mathbf{f}^C} + \mathbf \Theta \mathbf{t} = \mathbf{0} \quad \Leftrightarrow \quad \hat{\mathbf{t}} + \mathbf \Theta \mathbf{t} = \frac{\mathbf{n}}{\mathbf{f}^C},
\end{equation*}
the primal and dual feasibility constraints
\begin{equation*}
\mathbf{t} \geq 0, \quad \mathbf \Theta^\top \mathbf{f}^C \leq \mathbf{b},
\end{equation*}
the complementarity slackness condition
\begin{equation*}
\mathbf{t} \odot (\mathbf \Theta^\top \mathbf{f}^C - \mathbf{b}) = \mathbf{0}.
\end{equation*}
and objective's effective domain
\begin{equation*}
  \mathbf f^C \geq 0,
\end{equation*}
we obtain the same system of equations and inequalities as \eqref{eq:queue_system} up to the substitution of $f^U$ via \eqref{eq:lift_flow}.

Let us also derive the dual problem to \eqref{prob:queue_opt}:
\begin{equation*}
\max_{\mathbf t\geq 0} \min_{\mathbf{f}^C \geq 0} \quad \langle \hat{\mathbf{t}} + \mathbf \Theta \mathbf t, \mathbf{f}^C \rangle - \langle \mathbf{n}, \ln \mathbf{f}^C \rangle - \langle \mathbf t, \mathbf{b} \rangle.
\end{equation*}
Denoting $\mathbf s=\hat{\mathbf{t}} + \mathbf \Theta \mathbf t$ and finding that $\mathbf f^C(\mathbf t) = \frac{\mathbf n}{\mathbf{s}}$ this equals to
\begin{equation*}
\max_{\mathbf t \geq 0} \quad \langle \mathbf{s}, \frac{\mathbf{n}}{\mathbf{s}}  \rangle - \langle \mathbf{n}, \ln \mathbf{n} - \ln \mathbf{s} \rangle - \langle \mathbf t, \mathbf{b} \rangle. 
\end{equation*}
Simplifying the first term we get 
$\langle \mathbf{s}, \frac{\mathbf{n}}{\mathbf{s}}\rangle =\langle \mathbf{1}, {\mathbf{n}}\rangle = 1$ by $\mathbf{n} \in \Delta^C$.
Also $\langle \mathbf{n}, \ln \mathbf{n} \rangle$ does not depend on $\mathbf{t}$.
Thus, the dual problem up to the additive constant is
\begin{equation}\label{prob:queue_opt_dual}
  \max_{\mathbf t \geq \mathbf{0}} \quad \langle \mathbf n, \ln (\hat{ \mathbf t} + \mathbf \Theta \mathbf t) \rangle - \langle \mathbf{t}, \mathbf{b}\rangle.
\end{equation}

Since primal \eqref{prob:queue_opt} problem has strictly convex objective, its solution \( \mathbf{f}^C \) corresponding to any fixed distribution \( \mathbf{n} \) is unique.
A similar statement holds for the dual problem \eqref{prob:queue_opt_dual}: its objective is strictly concave on $\ker^\bot \mathbf \Theta$. 
Since changing $\mathbf t$ along $\ker \mathbf \Theta$ does not affect cycle waiting times $\mathbf \Theta \mathbf t$, 
this establishes a bijection between the distribution of users \( \mathbf{n} \) and the pair \( (\mathbf{f}^C,\mathbf \Theta \mathbf{t}) \) of cycle flows and cycle waiting times. 
% Consequently, the target function~--- such as the utility or the waiting time~--- can be represented purely as a function of \( \mathbf{n} \). 
%This functional dependence is central for reducing the problem to a variational inequality in terms of \( \mathbf{n} \) alone and simplifies both analysis and computational procedures.

Solving this system for $(\mathbf{f}^C, \mathbf \Theta \mathbf{t})$ using, e.g., an interior-point method provides the cycle waiting times necessary for evaluation of the cycles utilities $\mathbf{\boldsymbol \tau}(\mathbf{n})$ in the variational inequality framework and allows to obtain the distribution of user flows in an approximate solution to the equilibrium problem.

{\centering
  \section{Numerical experiments}\label{sec:exps}
}

Recall, that finding the equilibrium distribution \( \mathbf{n}^* \in \Delta^C \) is equivalent to variational inequality \eqref{eq:vi}:
\begin{align*}
    \langle \boldsymbol{\tau}(\mathbf{n}^*), \mathbf{n} - \mathbf{n}^* \rangle &\leq 0, \quad \forall \mathbf{n} \in \Delta^C.
\end{align*}

We solve the variational inequality using the Extragradient method \cite{korpelevich1977}, which iteratively updates the distribution by
\begin{align*}
    \mathbf{y}^{(k)} &= \Pi_{\Delta^C}\left( \mathbf{n}^{(k)} + \gamma \cdot \boldsymbol{\tau}(\mathbf{n}^{(k)}) \right), \\
    \mathbf{n}^{(k+1)} &= \Pi_{\Delta^C}\left( \mathbf{n}^{(k)} + \gamma \cdot \boldsymbol{\tau}(\mathbf{y}^{(k)}) \right),
\end{align*}
where \(\gamma > 0\) is the step size and \(\Pi_{\Delta^C}\) denotes the Euclidean projection onto the simplex \(\Delta^C\).

The Extragradient method is guaranteed to converge to the solution \( \mathbf{n}^* \) of the variational inequality under certain assumptions. Specifically, convergence holds if the mapping \( \boldsymbol{\tau}(\mathbf{n}) \) is monotone and Lipschitz continuous on the simplex \( \Delta^C \). Monotonicity ensures that for all \( \mathbf{n}, \mathbf{m} \in \Delta^C \),
\begin{align*}
    \langle \boldsymbol{\tau}(\mathbf{n}) - \boldsymbol{\tau}(\mathbf{m}), \mathbf{n} - \mathbf{m} \rangle &\le 0,
\end{align*}
and Lipschitz continuity requires the existence of a constant \( L > 0 \) such that
\begin{align*}
    \|\boldsymbol{\tau}(\mathbf{n}) - \boldsymbol{\tau}(\mathbf{m})\| &\le L \|\mathbf{n} - \mathbf{m}\|.
\end{align*}
for all \( \mathbf{n}, \mathbf{m} \in \Delta_C \). Under these assumptions, the Extragradient method converges at a rate of \( O(1/k) \) in terms of the squared Euclidean distance, i.e.,
\[
\|\mathbf{n}^{(k)} - \mathbf{n}^*\|^2 \leq \frac{C}{k}
\]
for some constant \( C > 0 \), to the unique equilibrium distribution \( \mathbf{n}^* \).

\begin{figure}[htpb]
  \centering
  \includegraphics[width=0.2\textwidth]{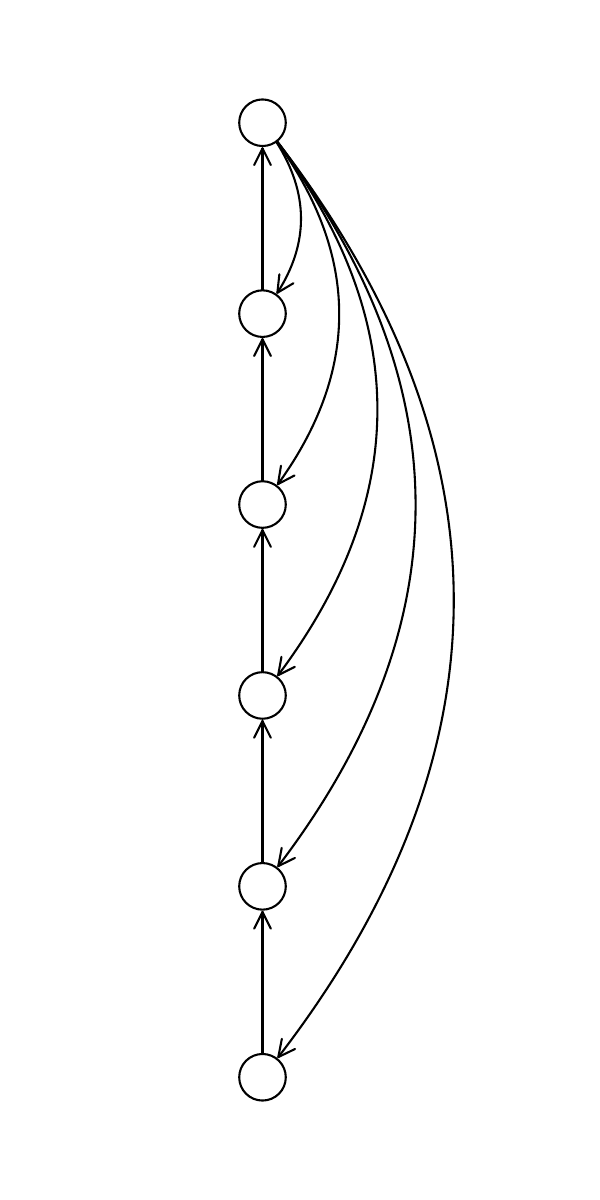}
  \caption{Synthetic network topology used in experiments}
  \label{fig:5lifts}
\end{figure}

We conducted experiments on a synthetic transportation network which is a larger variant of our toy network (Figure~\ref{fig:2lifts}) consisting of five cycles and five lifts, 
% where the membership matrix $\mathbf{\Theta}$ is upper-triangular,
see Figure~\ref{fig:5lifts}. 
We used CVXPY package \cite{cvxpy} for solving \eqref{prob:queue_opt} to find the cycle flows and cycle waiting times for given user distribution $\mathbf{n}$, as described in Section~\ref{sec:queue_opt}.
In our tests on the two-lifts toy network this implementation was always finding the solution to \eqref{eq:queue_system} with high accuracy.

The aim of the numerical experiment was to study the convergence behavior of the proposed equilibrium model under varying values of the attractiveness parameter \( a \).
Our results show that the model consistently converges to a stable equilibrium distribution \( \mathbf{n}^* \) for different parameter settings. To evaluate the quality of approximate solutions \( \mathbf{n} \), we use the variational inequality gap function defined as
\[
\operatorname{Gap}(\mathbf{n}) = \max_{\mathbf{z} \in \Delta^C} \langle \mathbf{\boldsymbol \tau}(\mathbf{n}), \mathbf{z} - \mathbf{n} \rangle \geq 0,
\]
which satisfies \( \operatorname{Gap}(\mathbf{n}^*) = 0 \) at the equilibrium.

\begin{figure}[htpb]
    \centering
    \includegraphics[width=0.8\textwidth]{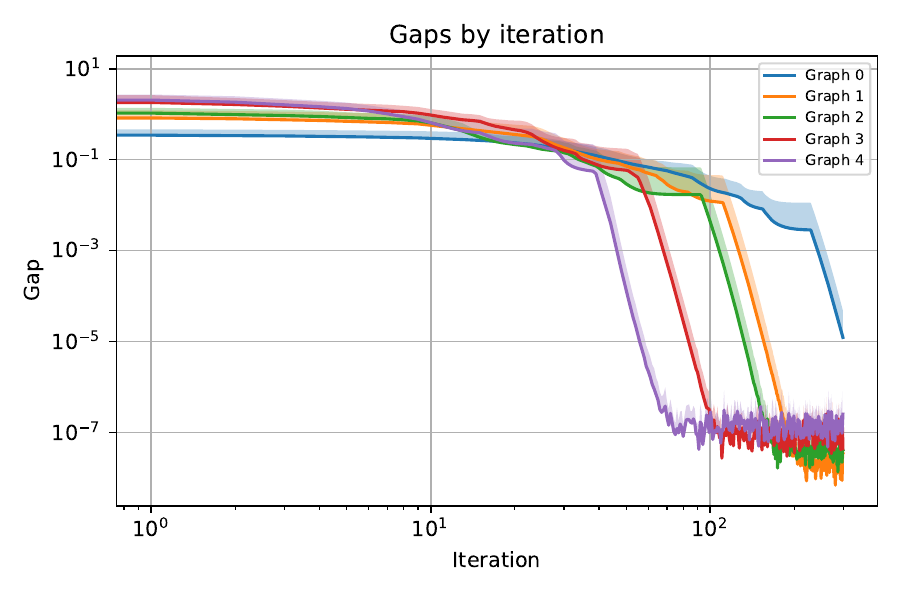}
    \caption{Convergence of the Extragradient method on a synthetic graph with five cycles and five elevators for different values of the attractiveness parameter \( a \). The plot shows the decrease of the variational inequality gap \( \operatorname{Gap}(\mathbf{n}) \) over iterations.}
    \label{fig:convergence}
\end{figure}

\begin{figure}[htpb]
    \centering
    \includegraphics[width=1.0\textwidth]{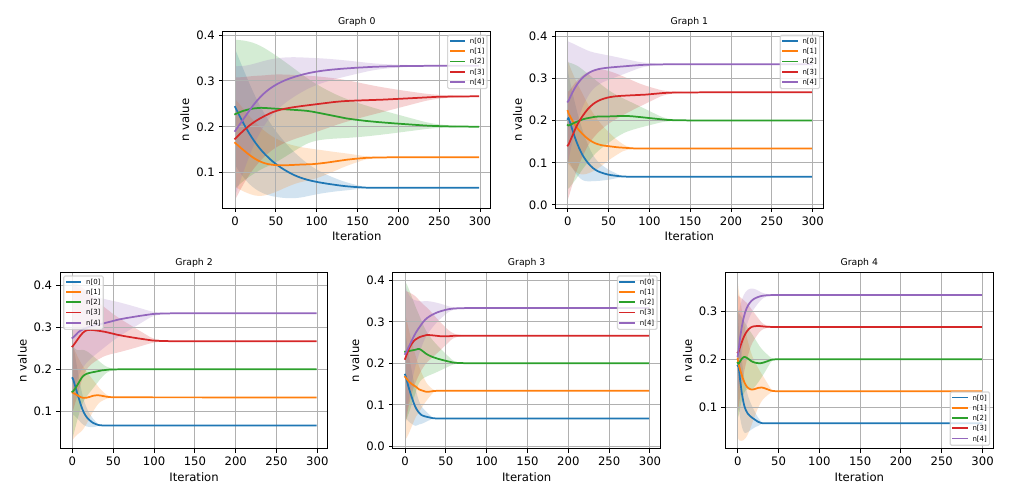}
    \caption{Convergence of distributions to the equilibrium \( \mathbf{n}^* \) from different random initializations. The plot illustrates that all trajectories approach the same equilibrium distribution, further confirming the robustness of the method and the good properties of the operator.}
    \label{fig:distribution_convergence}
\end{figure}

The numerical results illustrate a monotonic decrease of this gap over iterations, confirming the convergence and robustness of the Extragradient method across different attractiveness parameter values (see Figure~\ref{fig:convergence}).
Moreover, experiments with multiple random initializations demonstrate that not only the gap function decreases monotonically, but also the distributions themselves converge to the same solution.
As shown in Figure~\ref{fig:distribution_convergence}, trajectories starting from different initial points approach the same equilibrium distribution $\mathbf{n}^*$, indicating that the equilibrium is unique and that  operator $\boldsymbol \tau(\mathbf{n})$ exhibits favorable convergence properties.
This empirical evidence suggests that the operator is well-behaved and motivates further theoretical analysis of its properties.
All experimental code and results are available at \url{https://github.com/ThunderstormXX/Skymodeling}.

{\centering
  \section{Discussion}\label{sec:discussion}
}

One natural extension of the proposed equilibrium model involves a more structured definition of the cycle attractiveness parameters \( a_c \). Instead of assigning them arbitrarily, they can be formulated as a linear combination of interpretable features such as perceived cost and enjoyment:
\[
a_c = \alpha_1 \cdot \text{Costs}_c + \alpha_2 \cdot \text{Fun}_c,
\]
where \( \alpha_1 \) and \( \alpha_2 \) are tunable hyperparameters. This formulation allows for greater flexibility and interpretability in modeling user preferences and behavioral patterns in transportation systems.

More generally, it is reasonable to consider a high-dimensional feature space that includes many possible attributes of each cycle or route. In this setting, the attractiveness parameter \( a_c \) can be modeled as a function of a larger vector of features:
\[
a_c = f(\mathbf{x}_c),
\]
where \( \mathbf{x}_c \in \mathbb{R}^d \) denotes the feature vector for cycle \( c \). Methods from machine learning, such as principal component analysis (PCA), can then be employed to reduce the dimensionality of the feature space and identify the most influential directions of variation. This enables both interpretability and computational efficiency, while also opening the door for predictive modeling based on real-world data.

The use of dimensionality reduction techniques such as PCA allows one to extract the most informative components and visualize the trade-offs in attractiveness in a low-dimensional latent space. Moreover, it facilitates the integration of learned models into the equilibrium framework, providing a data-driven way to estimate preferences and simulate behavioral responses to changes in system parameters.

A deeper analysis of the mapping \( \boldsymbol \tau(\mathbf{n}) \), which reflects the interaction between the current distribution over cycles and congestion effects on elevators, is also important. Understanding its regularity and monotonicity properties is key to establishing theoretical guarantees such as convergence and uniqueness of the equilibrium solution. Although our current theoretical results rely on standard monotonicity and Lipschitz conditions for the cost operator \( \boldsymbol \tau(\mathbf{n}) \), empirical observations from synthetic experiments suggest behavior consistent with global convergence of the Extragradient method.

Once the parameters or models governing attractiveness are calibrated using data, the framework supports policy analysis. For instance, one may investigate how increasing penalties (represented by \( \text{Costs}_c \)) affects the equilibrium configuration. Since elevator congestion is endogenous to the equilibrium, it becomes possible to formulate optimization problems aimed at minimizing total congestion or delay by adjusting cost parameters, supporting practical applications in transportation system design.

{\centering
  \section{Conclusion}\label{sec:conclusion}
  In this work we proposed a novel flow-based ski resort model and found an efficient method for searching user equilibria in it.
  The findings suggest that this model could become a convenient tool for network optimization and management in ski resorts.
  A comparative study with real data is required to validate the model.
  Further research of problem properties is required to explain success of Extragradient method in solving the problem in the absence of operator monotonicity.
}

{\centering
\section*{Declarations}
}

\noindent
\textbf{Acknoledgments}
The authors thank the organizers of Traffic Flows on Networks Conference,
Sirius Mathematics Center, and Sirius University of Science and Technology that hosted the
Conference.

\vskip5pt

\noindent
\textbf{Funding}
The research is supported by the Ministry of Science and Higher Education of the Russian Federation, project No. FSMG-2024-0011

\noindent
\textbf{Data availability} This manuscript has no associated data.

\noindent
\textbf{Ethical Conduct} Not applicable.

\noindent
\textbf{Conflicts of interest}
The authors declare that there is
no conflict of interest.

\renewcommand{\refname}{\begin{center}{\Large\bf References} \end{center}}
\makeatletter
\renewcommand{\@biblabel}[1]{#1.\hfill}
\makeatother

\end{document}